\documentclass[dvips,12pt,a4paper]{article}
\usepackage{epsfig}
\usepackage{a4wide}
\usepackage{amssymb}
\begin{document}
\pagestyle{empty}
\vspace{1. truecm}
\begin{center}
\begin{Large}

{\bf Description of pp-interactions\\
with very high multiplicity at 70 GeV/c\\}

\end{Large}
\vspace{2.0cm} {\large  E. Kokoulina}
\\[0.3cm]
\small \sl The Gomel State Technical University, Belarus,
\end{center}

\begin{abstract}
\noindent The collective behavior of secondary particles in pp -
interactions at 70 GeV/c is  studied. A Two Stage Gluon Model is
offered to describe processes with very high multiplicity. An
active role of gluons is shown in multiparticle dynamics. The
analysis of multiplicity distributions has revealed a possibility
of a thermodynamic interpretation of these interactions. A
mechanism of the soft photon production as a signature of the
quark-gluon system is proposed.\\

 \vspace*{3.0mm} \noindent
\end{abstract}

These investigations have been carried out in the framework of the
project "Thermalization"(JINR). This project is aimed at studying
the collective behaviour of secondary particles in proton
interactions at 70 GeV/c \cite{SIS}.

According to the present understanding of hadronic structure,
based on quantum chromodynamics~(QCD)~\cite{QCD}, protons consist
of quarks and gluons. We have developed a Two Stage Gluon model
\cite{TSM} to describe high energy multiparticle production (MP)
in proton interactions. At the first stage of the model QCD and
thermodynamical approaches are used. At the second stage
(hadronization) a phenomenological description is applied.

After the inelastic collision of two protons a part of the energy
of these moving particles is converted into the thermal one.
Constituents of the proton, quarks and gluons, have enough energy
to be described by pQCD, because the strong coupling reaches a
small value. Quarks and gluons become liberated. Our model
investigations have shown that the quark division of primary
protons accompanied by hadron production in pp interactions at 70
GeV is absent. MP is realized by gluons. This study has confirmed
the idea of P.Carruthers about a passive role of quarks
\cite{CAR}.

We have used one of the most generally accepted methods to study
multiparticle dynamics: the multiplicity distribution (MD)
analysis. Two schemes of interactions are offered. They are
distinguished only by the quark-gluon (QG) stage. If we want to
study the gluon division inside the QG system (QGS), the scheme
branch model is used. If we are not interested in what is going
inside QGS the second scheme thermodynamic model should be
applied. In the both schemes some of gluons (not all of them)
leave QGS and convert to real hadrons. Using the thermodynamic
interpretation we can say that active gluons are evaporated from
hot QGS. After the evaporation they pass the stage of
hadronization.

Processes of pp interactions at 70~ GeV/c were investigated
experimentally \cite{BAB}. MD of charged particles were limited to
20 secondaries. Among them there are $n$ charged mesons ( $\pi^+$
or $\pi^-$) and two leading protons: $ p+p\rightarrow n\pi +2 N $.
In Thermalization project it is planned to observe events with
high multiplicities. The production of soft photons could give the
information about the early stage of QG interactions
\cite{SIS,SOF}.

The choice of the MP scheme is based on comparison with
experimental data~\cite{BAB}. Physicists from JINR (Dubna) used
generator PYTHIA and obtained MD of charged hadrons \cite{BAB}. It
was shown that PYTHIA generator does not agree with the
experimental data at high multiplicities and has a deviation two
orders of the magnitude at $n_{ch}=18$ equal to (Fig. 1). We have
refused from the quark model \cite{CHI} (Fig. 1) because of some
specific features which we cannot accept  and that is why we have
built a scheme of hadron interactions for MD description by means
of the  modified Two Stage Model (TSM) \cite{TSMA}. We consider
that at the early stage of pp interactions initial quarks and
gluons take part in the formation of QGS. They develop branch
processes \cite{GIO}. Also as in TSM we have used the hypothesis
of soft decoloration at the second stage
\begin{equation}
\label{2} P_n=\sum\limits_{m=0} P_m^PP_n^H(m),
\end{equation}
where $P_n$ - resulting MD of hadrons, $P_m^P$ - MD of partons
(quarks and gluons), $P_n^H(m)$~- MD of hadrons (the second stage)
from $m$ partons.

At the beginning of this study we used the model where some of
quarks from protons took part in the formation of hadron jets. But
it turned out that parameters of that model had values which were
differed a lot from the parameters obtained in $e^+e^-$-
annihilation, and especially from the parameters of hadronization.
It was one of the main reasons to give up the scheme with active
quarks.

We dwell upon the model where quarks of protons do not take part
in the jet production, but remain in the initial protons. All the
remained hadrons are formed by gluons. We  call these gluons the
active ones. It is important to know how many active gluons are in
QGS in the moment just after the impact.
 The simplest MD to describe this gluons
 is the Poisson distribution:

\begin{equation}
\label{4} P_k= e^{-\overline k} \overline k^k/k!,
\end{equation}
where $k$ and $\overline k$ are the number and mean multiplicity
of the active gluons, accordingly.

We begin our MD analysis with the branch scheme of gluons.  To
describe the MD of k gluons, we have used the Farry distribution
\cite{GIO}

\begin{equation}
\label{5} P_m^B= \frac{1}{\overline m^k}\left(1-\frac{1}{\overline
m} \right)^{m-k}\cdot \frac {(m-1)(m-2)\cdots (m-k+1)} {(k-1)!},
k>1,
\end{equation}
where m and $\overline m$ are the number of secondary gluons and
their mean multiplicites.

At the second stage some of the active gluons may leave QGS and
transform to real hadrons. We call such gluons the evaporated
ones. Let us introduce parameter $\alpha $ as the ratio of the
evaporated gluons, leaving QGS, to all the active gluons, which
may be transformed to hadrons. We use binomial distributions for
MD of the hadrons from the evaporated gluons at the stage of
hadronization

\begin{equation}
\label{9} P_n^H= C^{n-2}_{\alpha mN}\left(\frac{\overline n^h}
{N}\right)^{n-2}\left(1-\frac{\overline n^h} {N}\right)^{\alpha
mN-(n-2)},
\end{equation}
where $\overline n^h$ and $N$ are parameters of hadronization.
They mean the average value and maximal possible multiplicities of
the hadrons from one active gluon at the second stage. The hadron
MD in the process of pp scattering in the Two Stage Gluon Model
with branch(TSMB), is:

$$
P_n= \sum\limits_{k=0}\frac{e^ {-\overline k} \overline
k^k}{k!}\sum\limits_{m=k}\frac {1}{\overline m^k} \frac
{(m-1)(m-2)\dots(m-k+1)}{(k-1)!}
$$

\begin{equation}
\label{12}
\cdot \left(1-\frac{1}{
\overline m}\right)^{m-k}
C^{n-2}_{\alpha mN}\left(\frac{\overline n^h}
{N}\right)^{n-2}\left(1-\frac{\overline n^h}
{N}\right)^{\alpha mN-(n-2)}.
\end{equation}
The parameters are determined from the comparison with
experimental data \cite{BAB}. They are $N=40$(fix), $\overline
m=2.61 \pm .08$, $\alpha = .47 \pm .01$, $\overline k=2.53 \pm
.05$, $\overline n^h=2.50 \pm .29$. We can conclude that branch
processes are absent, since parameters $\overline m$ and
$\overline k$ are equal to the errors. The fraction of the
evaporated gluons is equal to .47. A maximal possible number of
hadrons from the gluon looks very much like the number of partons
in the glob of cold QG plasma of L.Van Hove \cite{LVH}. At the
fixed parameter of hadronization $\overline n^h$ equal to $1.63$
(see below the thermodynamic model) the fraction of the evaporated
gluons is about .73 (Fig. 1).

In the thermodynamic model without branches the active gluons
which appeared in the moment of the impact, may leave QGS and
fragment to hadron jets. We assume that the active gluons the
evaporated from QGS have the Poisson MD as in (\ref{4}).
 Using the binomial distribution for hadrons
(\ref{9}) and the idea of convolution of two stages (\ref{2}), we
obtain MD of hadrons in pp-collisions in the framework of the Two
Stage Thermodynamic Model (TSTM)

\begin{equation}
\label{13} P_n=\sum\limits_{m=0}^{M}\frac{e^{-\overline m}
\overline m^m}{m!} C^{n-2}_{mN}\left(\frac{\overline n^h}
{N}\right)^{n-2}\left(1-\frac{\overline n^h} {N}\right)^{mN-(n-2)}
(n>2)
\end{equation}
($P_2=e^{-\overline m}$). The comparison (\ref{13}) with
experimental data \cite{BAB}\quad (see Fig. 2), gives the
following parameter values: $N=4.24 \pm .13$, $\overline m=2.48
\pm .20$, $\overline n^h=1.63\pm .12$. In sum~(\ref{13}) we
constrain the maximal possible number of the evaporated gluons
equal to $M=6$. At the description of experimental data of
$e^+e^-$ annihilation the hadronization parameter $N$ was found
equal to $\sim 4-5$ \cite{TSMA}. We can see that our parameter $N$
obtained in TSTM coincides with this value. Both TSMB and TSTM
describe the data equally well.

The maximal possible number of the charged particles from TSTM is
$26$. It is interesting to get MD for neutral mesons. For this
purpose we have taken the $\pi ^0$ mean multiplicity in
pp-interactions at 69 GeV/c and used it for normalization. It is
equal to $2.57\pm .13$ \cite{MUR}. We have determined the
parameter of hadronization $\overline n^h_0$ for neutral mesons.
It is equal to $1.036 \pm .041$. We are based on TSM \cite{TSM} to
estimate the probability of the different hadron production from
one active gluon.

MD for neutral mesons is shown in Fig. 3. From this distribution
we see that the maximal possible number of $\pi ^0$'s from TSTM is
equal to 16. MD for total multiplicity is shown on Fig. 4. We see
that the maximal possible number of total number of particles in
this case is equal to42.

The dependence of the mean multiplicities of neutral mesons versus
the number of the charged particles may be obtained by means of MD
for total multiplicity $P_{n_{tot}}$:
\begin{equation}
\label{15} \overline n_0(n_{ch})=\sum \limits _{n=n_1}^{n_2}
P_{n_{tot}} \cdot (n-n_{ch})/\sum \limits _{n=n_1}^{n_2}
P_{n_{tot}},
\end{equation}
we take the Bayess theorem into account, $n_1$ and $n_2$ are
determined only by conservation laws. Fig. 5 shows a difference
with the data at small multiplicities. The noticeable improvement
will be reached if we decrease the top limit at low multiplicities
($n_{ch}\leq 10$) to $n_2=2 n_{ch}$ (Fig. 6). We don't know what
is happening in the region of VHMP. This behavior of $n_1$ and
$n_2$ in (\ref{15}) indicates that Centauro events with a large
number of charged particles and practically no accompanying
neutrals, may be realized in the region of VHMP. AntiCentauro
events with a large number of neutral particles and with very
small number charged ones must be absent.

In Two Stage Model with gluon branch it was established that
several of active gluons are staying inside of hot QGS and don't
give hadron jets. New formed hadrons are catching up small
energetic gluons which were free before this time. These hadrons
are excited because they have additional energy at the expense of
absorbed gluons. This energy may be thrown down by means of the
photon radiation.

In project "Thermalization" it is planned to investigate of soft
photons (SP) \cite{SIS}. It was shown that measured cross sections
of such photons are several times larger than expected from QED.
For the explanation of the SP excess the phenomenological glob
model \cite{LVH} and the modified soft annihilation model Lichard
and Thomson \cite{SAM} are used.

We want to estimate the number of them. We consider that at
certain moment QGS or exited hadrons may be in almost equilibrium
state on. That's why we try to use for the description of the
massless bosons (gluons and photons) the black body emission
spectrum \cite{HER}

\begin{equation}
\label{16}
\frac {d \rho (\nu )}{d \nu }=
\frac {8\pi }{c^3} \frac {\nu ^2}{e^{h\nu /T}-1},
\end{equation}
where $ \nu $ is the energy of photon. These spectra help us to
calculate the number of SP \cite{KUR}. The gluon density at the
deconfinement temperature $T_c\approx 160-200$ MeV can be
estimated by comparison with relic one: $\rho_{gl}(160)=0.13
(fm)^{-3}$ and $\rho _{gl}(200)=0.25 (fm)^{-3}$. The number of
gluons $N_{gl}$ in the hot QGS of size $\sim L^3$, where $L=2$fm,
will be: $N_{gl}(160)\sim 1$, $N_{gl}(200)\sim 2$.

Using the spectral spatial density of relic photons (\ref{16}) it
is possible to get the number of SP $N_\gamma $ in the region of
size of our system (new formed hadrons). This size must be bigger
than one in the gluon case. If the size of our system about 2 fm
and average energy of photons 15-20~MeV/c the number of such SP
will be the order of $10^{-3}$.

I thank members of Organizing Committee for wonderful possibility
to take part in ISMD XXXIII. I also are deeply indebted to
V.Nikitin for considerable support of my study. I thank E.Kuraev
for the help in the understanding of SP nature.

\newpage

\begin{figure}
\begin{minipage}[b]{.3\linewidth}
\centering
\includegraphics[width=\linewidth, height=2in, angle=0]{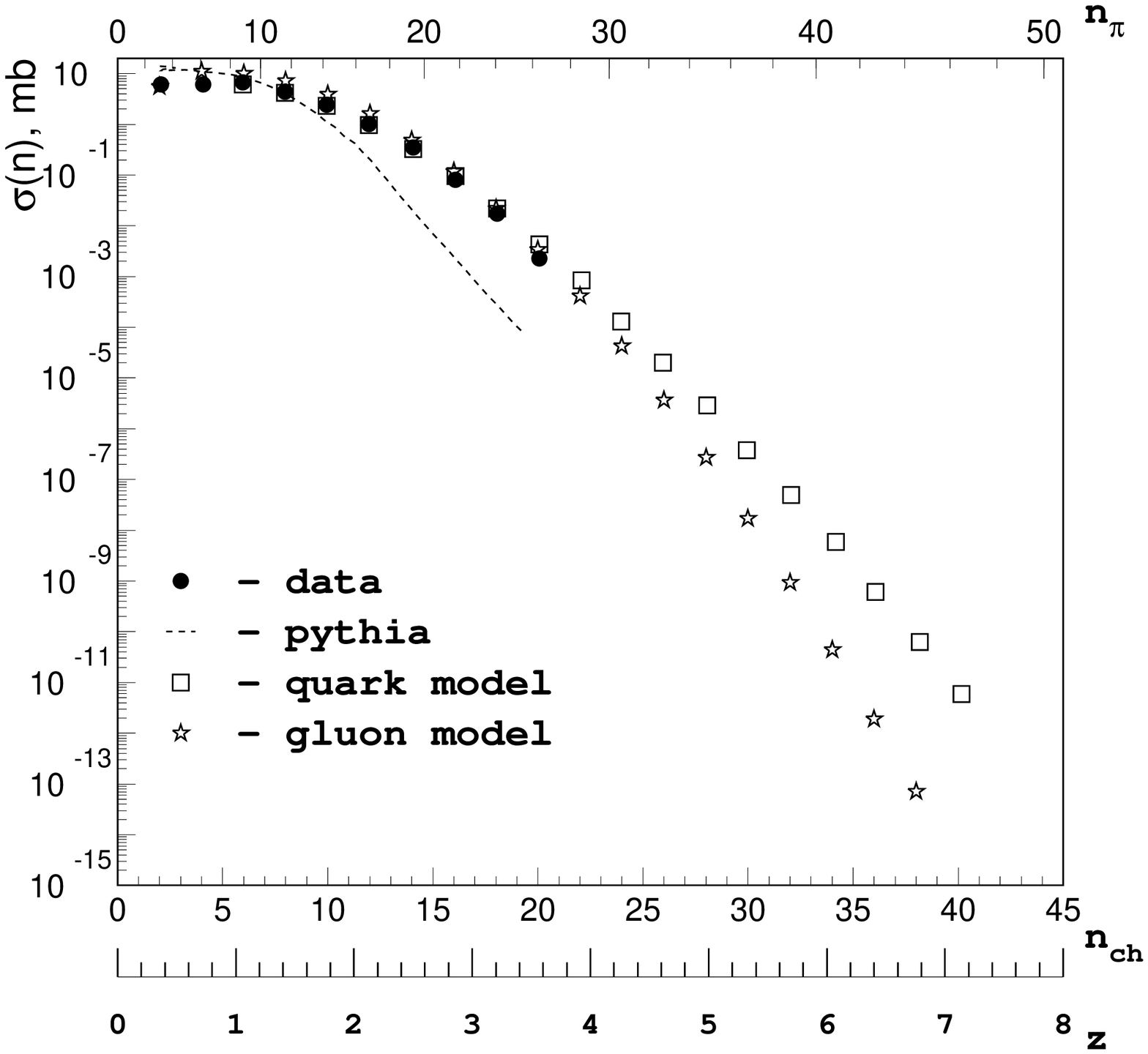}
\caption{$\sigma (n_{ch})$ in different models (see text).}
\label{1dfig}
\end{minipage}\hfill
\begin{minipage}[b]{.3\linewidth}
\centering
\includegraphics[width=\linewidth, height=2in, angle=0]{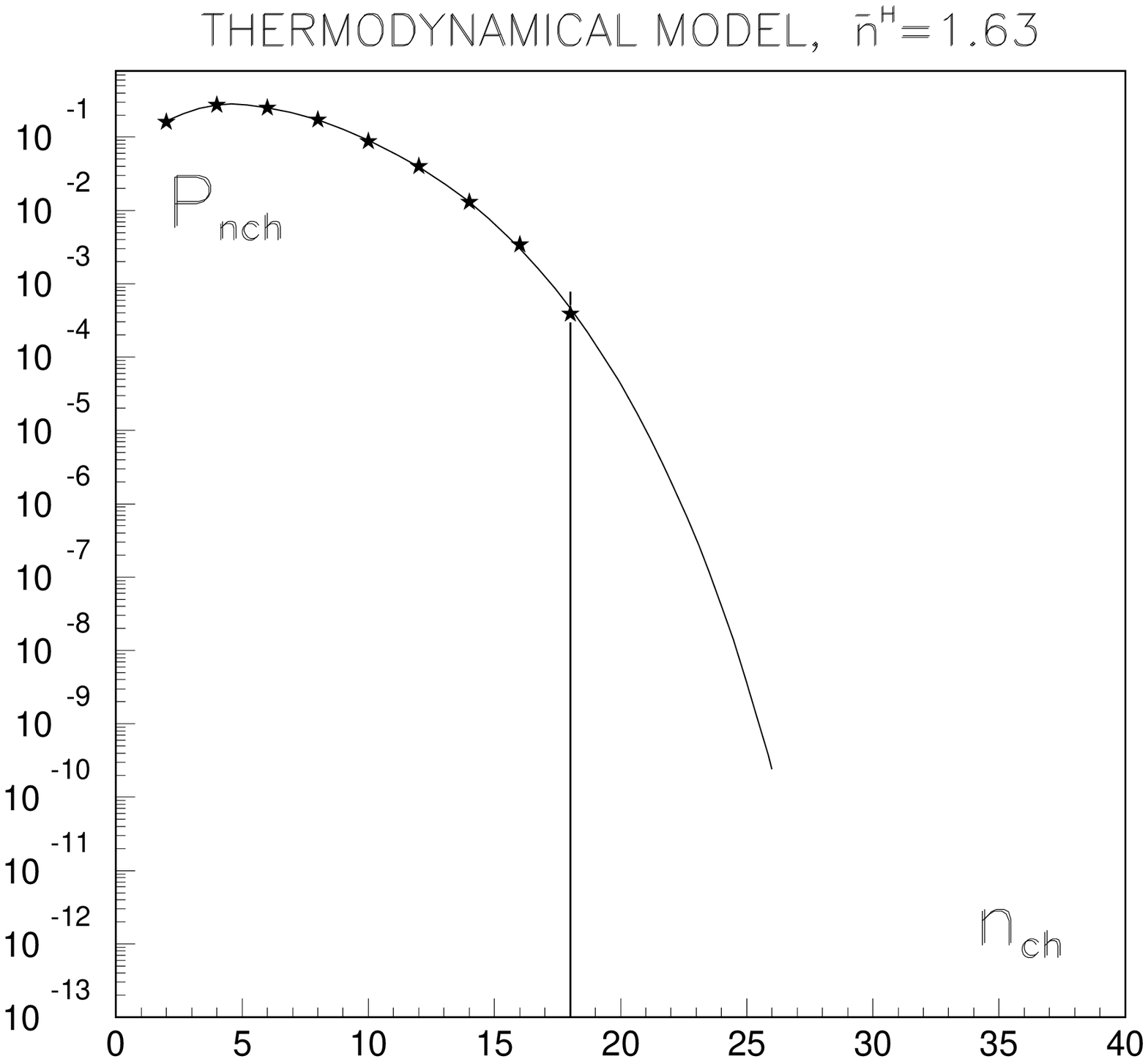}
\caption{MD $P(n_{ch})$ in TSTM.}
\label{2dfig}
\end{minipage}\hfill
\begin{minipage}[b]{.3\linewidth}
\centering
\includegraphics[width=\linewidth, height=2in, angle=0]{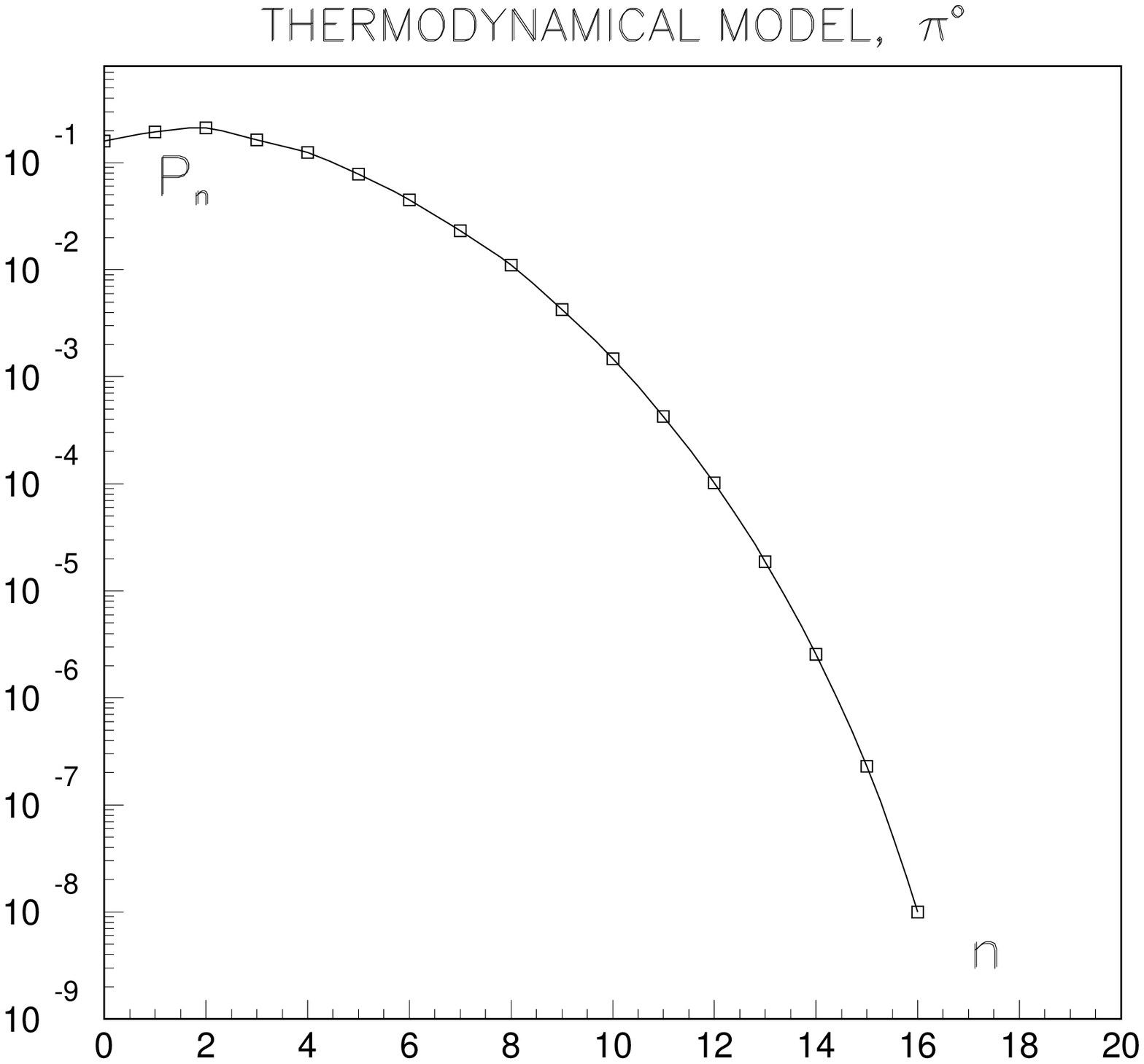}
\caption{MD $P(n_o)$ in TSTM.}
\label{3dfig}
\end{minipage}
\end{figure}

\begin{figure}
\begin{minipage}[b]{.3\linewidth}
\centering
\includegraphics[width=\linewidth, height=2in, angle=0]{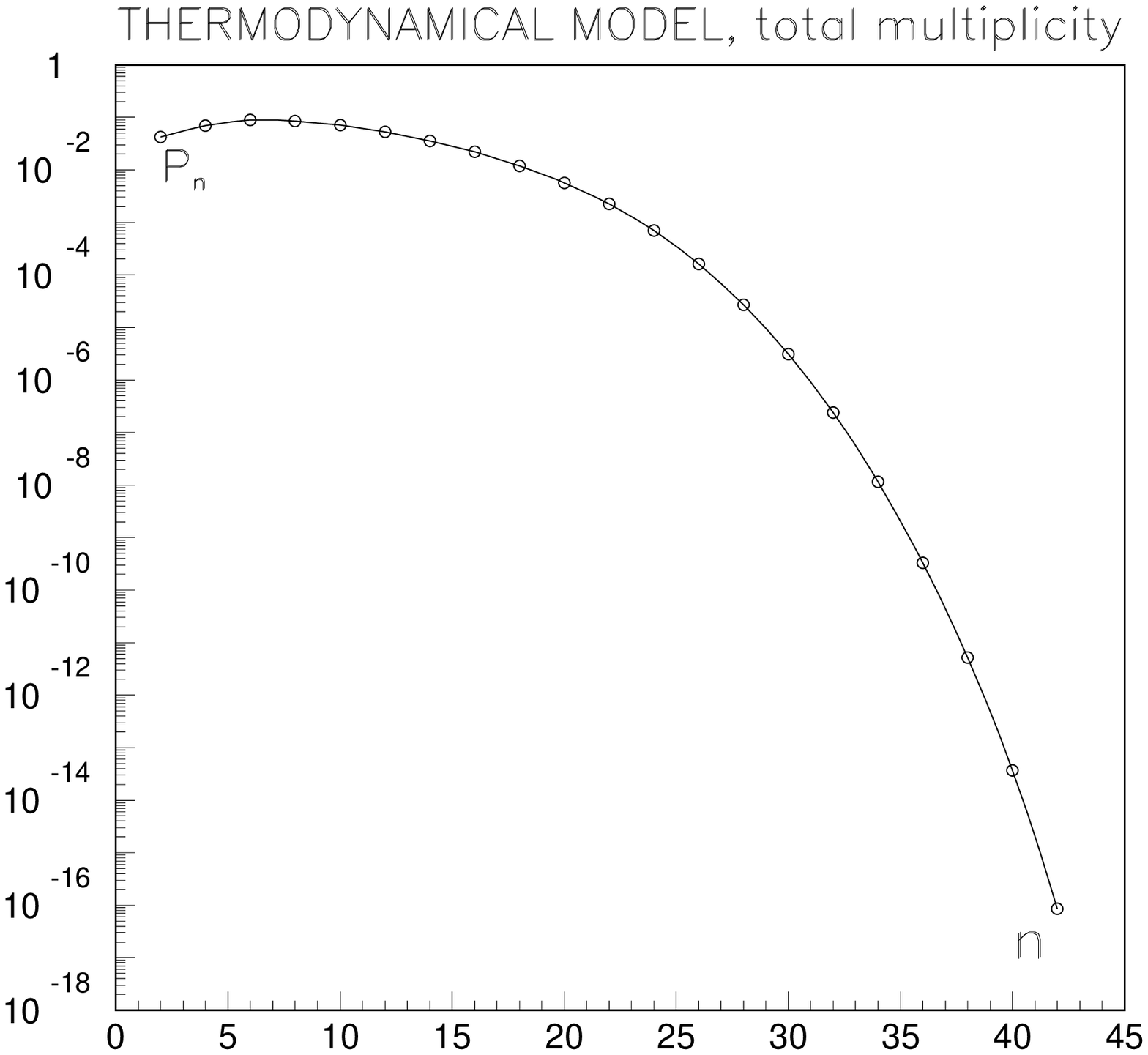}
\caption{MD $P(n_{tot})$ in TSTM}
\label{4dfig}
\end{minipage}\hfill
\begin{minipage}[b]{.3\linewidth}
\centering
\includegraphics[width=\linewidth, height=2in, angle=0]{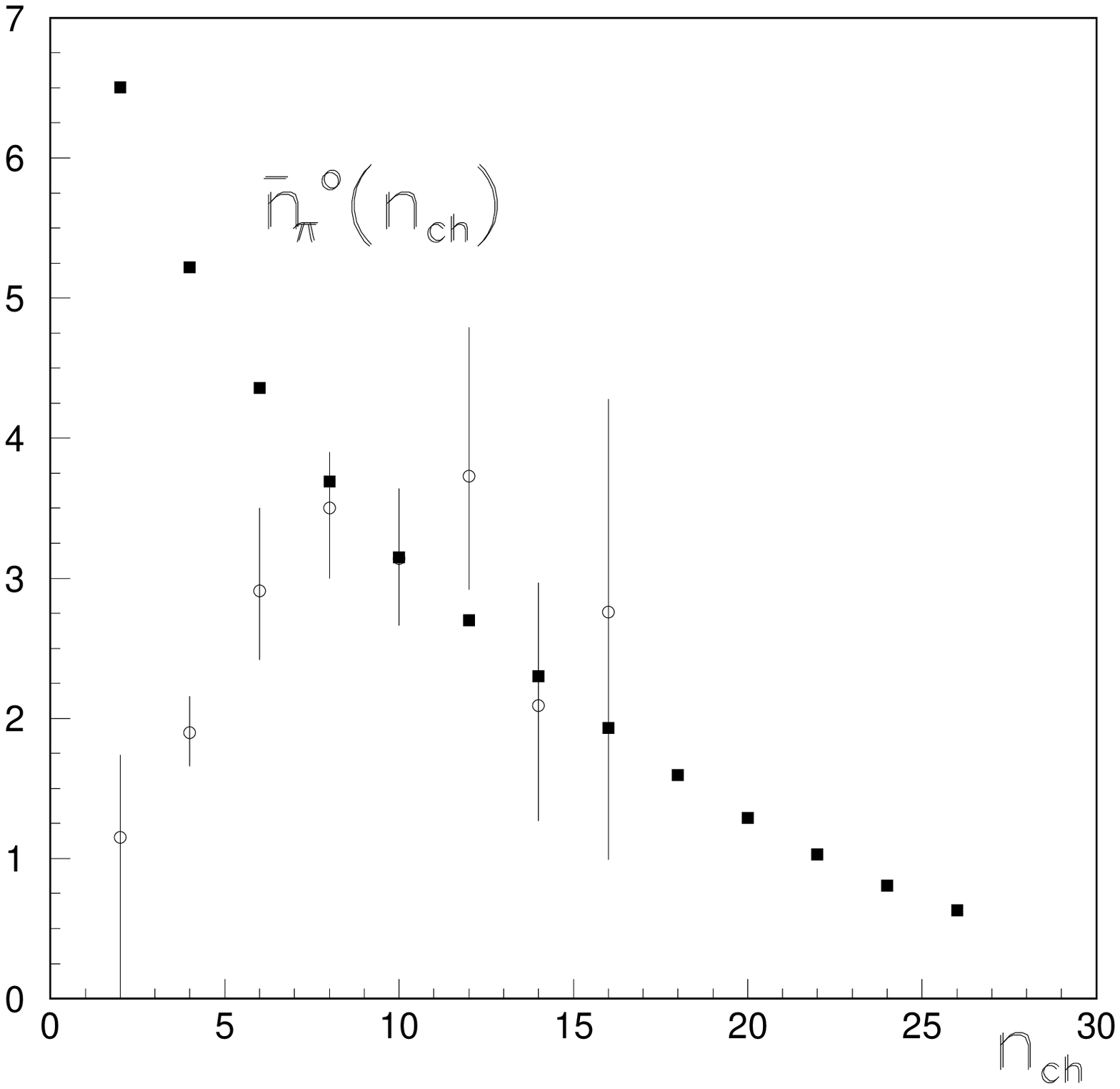}
\caption{$\overline n_\pi^o$ versus $n_{ch}$ without of
restrictions.} \label{5dfig}
\end{minipage}\hfill
\begin{minipage}[b]{.3\linewidth}
\centering
\includegraphics[width=\linewidth, height=2in, angle=0]{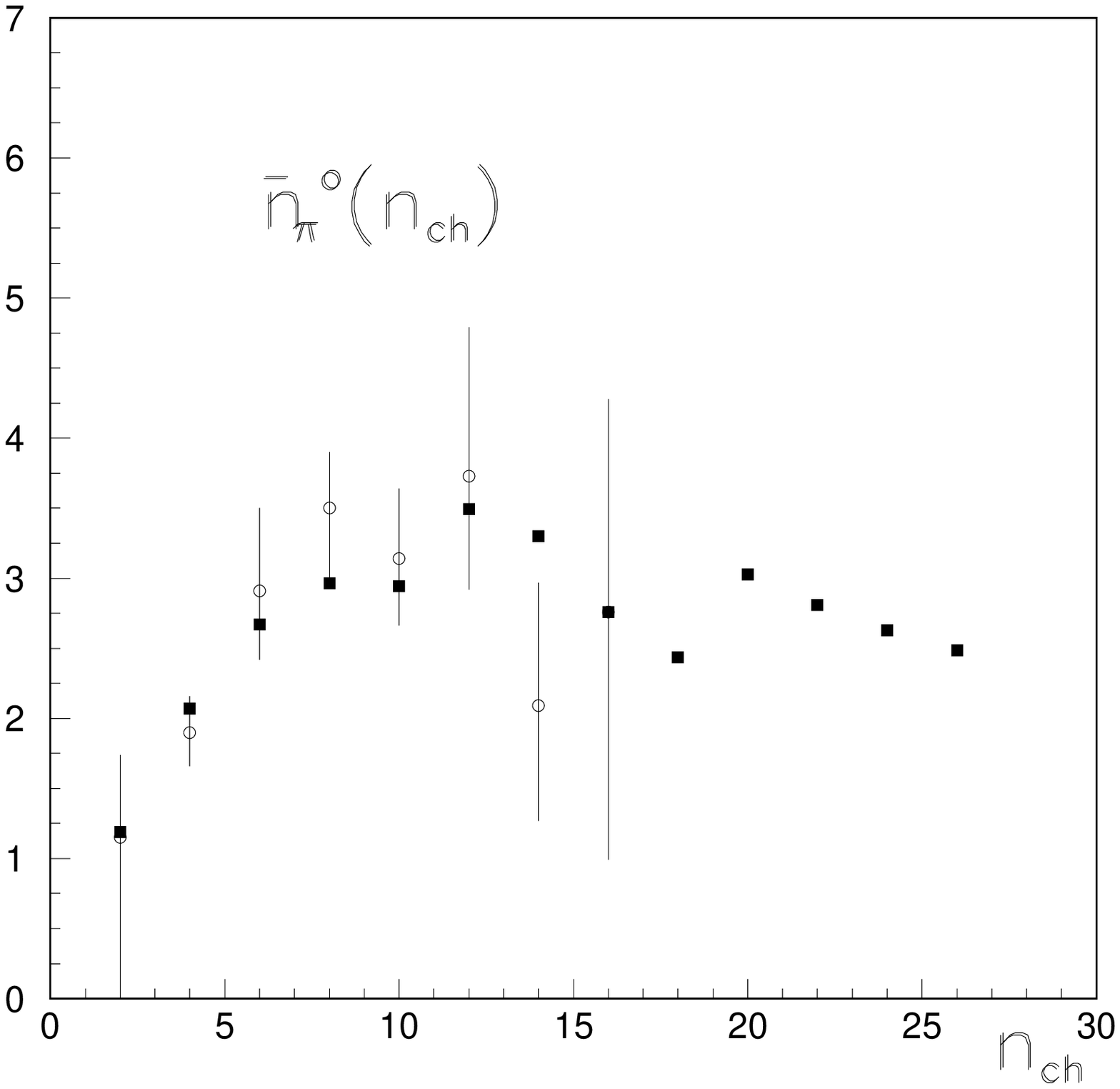}
\caption{$\overline n_\pi^o$ versus $n_{ch}$ with restrictions.}
\label{6dfig}
\end{minipage}
\end{figure}

\end{document}